\documentclass[superscriptaddress,twocolumn,
amssymb,amsmath,nobibnotes,aps,pre,showkeys,showpacs]{revtex4-1}
\usepackage{bm}
\usepackage{bbm}
\usepackage{latexsym}
\usepackage{dcolumn}
\usepackage{amsfonts,amssymb,amsmath}
\usepackage{graphicx,epsfig}
\usepackage{psfrag}
\usepackage{subfigure}
\usepackage{color}
\usepackage{ulem}
\newcommand{\hh}{\widehat{H}}
\newcommand{\vv}{\widehat{V}}
\newcommand{\ff}{\widehat{F}}
\newcommand{\uu}{\widehat{U}}
\newcommand{\uue}{\widehat{\mathcal U}}
\newcommand{\jj}{\widehat{J}}
\newcommand{\fl}{\widehat{\mathcal F}}
\newcommand{\be}{\begin{equation}}
\newcommand{\ee}{\end{equation}}
\newcommand{\hht}{\widehat{\mathcal{H}}}
\newcommand{\vvt}{\widehat{\mathcal{V}}}

\begin{document}

\title{Self-similar spectrum in effective time independent Hamiltonians for kicked systems}

\author{Rashmi Jangid Sharma}\email{jangid.rashmi@gmail.com}
\affiliation{Department of Physics, Birla Institute of Technology and Science, Pilani 333031, India.} \author{Jayendra N. Bandyopadhyay}\email{jnbandyo@gmail.com} \affiliation{Department of Physics, Birla Institute of Technology and Science, Pilani 333031, India.} \author{Tapomoy Guha Sarkar}\email{tapomoy1@gmail.com}
\affiliation{Department of Physics, Birla Institute of Technology and Science, Pilani 333031, India.}

\begin{abstract}
We study multifractal properties in the spectrum of effective time-independent Hamiltonians obtained using a perturbative method for a class of delta-kicked systems. The evolution operator in the time-dependent problem is factorized into an initial kick, an evolution dictated by a time-independent Hamiltonian, and a final kick. We have used the double kicked $SU(2)$ system and the kicked Harper model to study butterfly spectrum in the corresponding effective Hamiltonians. We have obtained a generic class of $SU(2)$ Hamiltonians showing self-similar spectrum. The statistics of the generalized fractal dimension is studied for a quantitative characterization of the spectra. 
\end{abstract}

\pacs{05.45.Df, 05.45.Mt}

\maketitle

Hamiltonian systems undergoing periodic delta kicks are studied extensively as a generic model for classical and quantum chaos \cite{qchaos}. This has found new relevance in the possibilities of engineering such systems using ultra-cold atoms \cite{cold_atom}. In the traditional approach, such time-dependent systems are theoretically studied using the Floquet analysis whereby the quasienergy spectrum is investigated for the signature of quantum chaos \cite{qchaos} and quantum criticality with varying parameters of the Hamiltonian \cite{demolish,our_paper}. Quantum chaos studies have also shown the existence of fractal butterfly patterns in the quasienergy spectrum of periodically driven systems \cite{khm,gong1,gong2,gong3} indicating an infinite number of quantum phase transitions \cite{goldman}. These systems are particularly interesting in the fact that though their classical phase space dynamics may be chaotic, the quantum quasienergy spectrum does not follow the celebrated Bohigas-Giannoni-Schmit conjecture \cite{bohigas}. We focus on two such systems. The first one consisting of driven $SU(2)$ operators is also known as the double kicked top model. This system is of interest as it finds realization through driven two-mode BEC systems \cite{gong2,gong3}. Secondly, we look at the kicked Harper model. The time-independent version of the Harper model represents the behavior of electrons in periodic lattice in the presence of a constant magnetic field in the tight-binding nearest neighbor approximation \cite{harper,hoftstadter}.

It is possible to construct a time-independent effective approximate Hamiltonian for such time-dependent systems when the frequency of the periodic driving is large. Traditionally, the effective Hamiltonian is obtained from the Floquet operator using the Cambell-Baker-Hausdorff (CBH) or Trotter expansion. It has been shown that the CBH method to study the kicked systems suffer intrinsic flaws and an alternative formulation \cite{fishman,dalibard} is better suited for more accurate analysis of such systems \cite{our_paper}. The effective Hamiltonian thus obtained is found to mimic the exact time-evolution for a large range of parameter values.

In this paper, we have investigated for fractal spectrum in the effective time-independent Hamiltonian obtained from the Floquet operator using the perturbative method used in earlier works \cite{fishman,dalibard}. We have used the double kicked $SU(2)$ system \cite{gong2,gong3} and the kicked Harper model \cite{khm} to study fractal spectrum in the corresponding effective Hamiltonians. We have obtained a generic class of $SU(2)$ Hamiltonians showing self-similar spectrum in finite dimension. These models revealing butterfly spectra also contain the Harper model as a special case. We study in detail the (multi)fractal properties of the eigenvalues and eigenstates of the effective Hamiltonians for all the models considered. The statistics of the generalized fractal dimension is studied to quantitatively understand their scaling behavior.


A general time-dependent problem where $\hh(t) = \hh_0 + \vv(t)$, with a time-periodic potential $\vv(t)=\vv(t+T)$ of periodicity $T$ has a Floquet operator $\fl(t)$ which corresponds to the time-evolution operator for one time-period. The traditional approach to extract an effective static Hamiltonian one writes $\fl = \exp(- i \hh_{\rm eff} T )$ and uses the CBH expansion to read out $\hh_{\rm eff}$ up to any order in $T \sim 1/\omega$. This method, however has been found to suffer from several inadequacies. 

The method used in Refs. \cite{fishman,dalibard} expresses the time-evolution operator $\uu(t_i \rightarrow t_f)$ between time instants $t_i$ and $t_f = t_i + T$, as a sequence of operations consisting of an initial kick followed by an evolution under a time-independent Hamiltonian and a final `micro-motion' 
\be 
\uu(t_i \rightarrow t_f) =
\uue^\dagger(t_f) e^{-i \hh_{\rm eff} T}\, \uue(t_i)
\label{eq:floquet}
\ee
where $\uue(t) = e^{i \ff(t)}$ such that $\ff(t) = \ff(t+T)$ with zero average over one time period. For high-frequency pulsing, the operators $\hh_{\rm eff}$ and $\ff(t)$ can be expanded as a perturbation series in $1/\omega$ of the form 
\be
\hh_{\rm eff} = \sum\limits_{n = 0}^\infty\, \frac{1}{\omega^n} \hh_{\rm eff}^{(n)},~~~~\ff(t) = \sum\limits_{n = 1}^\infty\, \frac{1}{\omega^n} \ff^{(n)}.
\ee 
This ansatz along with Eq. \eqref{eq:floquet} can be used to obtain 
$\hh_{\rm eff}$ and $\ff(t)$ up to any desired accuracy. In this method, the average time-independent part is retained in $\hh_{\rm eff}$ and all the time-dependence is pushed to the operator $\ff(t)$ at each order of perturbation. The convergence of the perturbation series has been surmised in earlier works \cite{fishman,dalibard}. The periodic potential $\vv(t)$ may be expanded in a Fourier series as $\vv(t) = \vv_0 + \sum\limits_{n=1}^\infty \, \Bigl(\vv_{n} e^{i n \omega t} + \vv_{-n} e^{- i n \omega t}\Bigr)$.
The truncated series for $\hh_{\rm eff}$ and $\ff(t)$ up to $\mathcal{O}(1/\omega^2)$ can be expressed in terms of the Fourier coefficients of $\vv(t)$ \cite{dalibard,supplement}. We use this as the general expression for the effective Hamiltonian for periodically driven systems. 

\begin{figure}[t]
\includegraphics[height=7.0cm,width=8cm]{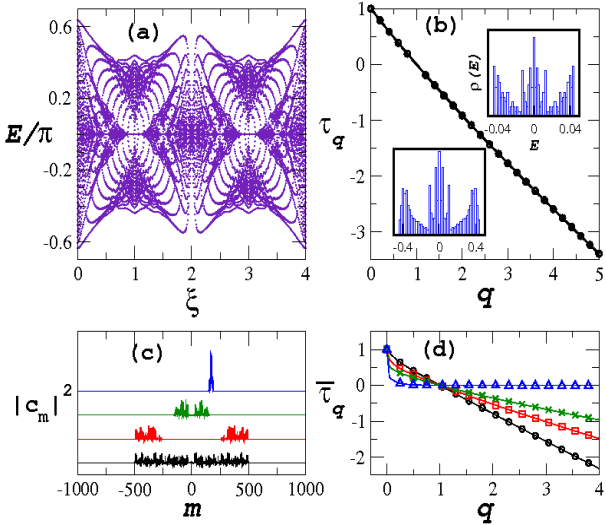}
\caption{(Color online) (a) Folded energy spectrum of the effective time-independent Hamiltonian of the double kicked top showing butterfly pattern. (b) Multifractal scaling exponent showing a linear dependence on $q$ for $\eta/j = G_r =(\sqrt{5}-1)/2$. Insets: Self-similar distribution of energy spectrum on different scales. (c) Eigenvectors in $\jj_z$ eigenbasis. $D_2$ values are $0.002$, $0.218$, $0.511$, and $0.855$ from top to bottom. Localized states have smaller $D_2$. (d) Corresponding multifractal behavior for the same eigenstates. Slopes in the linear region are $-0.003\, (\triangle)$, $-0.314\, (\times)$, $-0.489\, (\square)$, and $-0.771\, (\circ)$.}
\label{fig:fig1}
\end{figure}

We consider the double kicked top model with the Hamiltonian \cite{gong2,gong3}
\be 
\hh = \frac{2\alpha}{T} \jj_x + \frac{\eta}{2 j} \jj_z^2 \sum_{n=-\infty}^{+\infty} \left[ \delta\left(t-nT-\frac{T}{2}\right) - \delta(t-nT) \right].
\ee
The $\jj_i$s here represent $SU(2)$ generators in the $d = (2j+1)$ dimensional Hilbert space. The corresponding Floquet operator is given by \cite{gong2,gong3}
\be
\fl = \exp{\left\{-i \alpha \jj_+ e^{i[\eta(2\jj_z + \mathbbm{1})/2j]} + {\rm h.c.}\right\}} \exp\left(- i \alpha\jj_x\right),
\label{eq:fl_dkt}
\ee
where $\jj_+$ denotes the operator $(\jj_x + i \jj_y)/2$. The quasiperiodic nature of the factor $e^{i[\eta(2\jj_z+\mathbbm{1})/2j]}$ for irrational values of the parameter $\eta/j$ leads to interesting spectral properties \cite{gong2,gong3}. The above Floquet operator can also be obtained from a different driven $SU(2)$ Hamiltonian of the form 
\be
\begin{split}
\hht &= \hht_0 + \vvt \sum\limits_{n=-\infty}^{+\infty} \delta(t-nT), ~~~{\rm where}~~\vvt = \alpha \jj_x \\
{\rm and}&~~~\hht_0 = \alpha \frac{\jj_+}{2 T} \exp\left[i \frac{\eta}{2j}\left(2\jj_z+ \mathbbm{1}\right) \right] + {\rm h.c.}
\end{split}
\ee
This is a single kicked system whose Floquet operator, given in Eq. \eqref{eq:fl_dkt}, matches exactly with that of the double kicked top and thereby exhibits interesting Cantor set properties in the quasienergies spectrum. The possibility of experimental realization of this system has been studied \cite{haake}. We are interested in the spectral properties of the effective approximate static Hamiltonian corresponding to this system. The effective Hamiltonian is given by
\be
\begin{split}
\hht_{\rm eff} &= \hht_0 + \frac{\vvt}{T} + \frac{1}{\omega^2 T^2} \bigl[\bigl[\vvt, \hht_0], \vvt\bigr] \left(\sum\limits_{n=1}^\infty \frac{1}{n^2}\right)\\ &= \hht_0 + \frac{\vvt}{T} + \frac{1}{24} \bigl[\bigl[\vvt, \hht_0], \vvt\bigr]
\end{split}
\label{eq:heff_dkt}
\ee 

\begin{figure}[b]
\includegraphics[height=5cm,width=8cm]{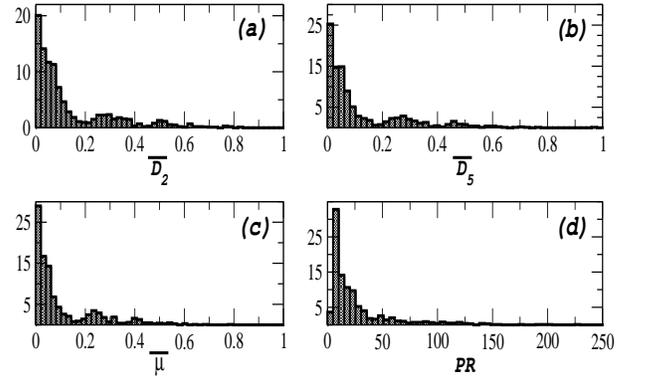}
\caption{(a)-(b): The distribution of $\overline{D}_2$ and $\overline{D}_5$ for all the $2j+1=2001$ eigenstates. (c) The distribution of the slope $\mu$ of the $\overline{\tau}_q$ vs. $q$ curves in the linear region. (d) The distribution of the participation ratio indicating that the bulk of the eigenstates are localized.}
\label{fig:fig2}
\end{figure}

Fig. \ref{fig:fig1}(a) shows the folded energy spectrum of the Hamiltonian $\hht_{\rm eff}$ as a function of $\xi = \eta/\pi j$ for $\alpha = 1/j$ where we have chosen an even value of spin $j=20$. We note that odd values of $j$ would bring about changes in the spectrum. The spectacular butterfly appearance for the static approximate eigenspectrum is in remarkable agreement with the quasienergy spectrum of the original double kicked top \cite{gong2,gong3}. The spectrum shows qualitative similarity with the Hoftstadter butterfly \cite{hoftstadter} owing to the presence of the quasiperiodic term. This feature is however along the off-diagonal nearest neighbor band and is therefore different from the Harper/Hofstadter case where it appears along the diagonal. In order to study multifractality of the energy spectrum, we set $\eta/j$ at an irrational value of the golden ratio $G_r=(\sqrt{5}-1)/2$ and a large value of $j=2500$. To study the statistical property of the energy spectrum, we consider histogram of eigenvalues for different scales. The level distribution $\rho(E)$ exhibits remarkable self-similarity as seen in insets of Fig. \ref{fig:fig1}(b). The quantitative measure of the self-similar behavior is  done by using the generalized fractal dimension. Dividing the full range of the energies into $N$ bins of size $s$ each we use the standard box-counting to obtain the probability $p_s(i)$ of finding a given energy eigenvalue in the $i$-th bin. The scaling exponent $\tau_q$ is related to the $q$-th moment via the partition function $Z_q(s) = \sum\limits_{i=1}^N p_s(i)^{q} \sim s^{\tau_q}$. The generalized fractal dimension is defined as $D_q = \tau_q/(1-q))$ \cite{stanley}. Figure \ref{fig:fig1}(b) shows that for large values of $j$ the spectrum indeed shows multifractal behavior with $\tau_q$ linearly decreasing with increasing $q$ with a slope $-0.871$. The quantitative measure $D_2$ for this spectrum has the value $0.913$. To look for fractal behavior in the eigenvectors of $\hht$, we consider four eigenvectors shown in Fig. \ref{fig:fig1}(c) with different localization properties. The highly localized eigenvectors manifest as the ones which have very sharp support over a narrow band of the index $m$ labeling the components in $\jj_z$ eigenbasis. To define the scaling exponent $\bar{\tau}_q$ for the $n$-th eigenvector we consider its components $\{c_m^{(n)}\}$ and divide the total dimension $d=2j+1$ into $M$ partitions, and define $\tilde{p}(i) =\sum |c_m^{(n)}|^2$ where the summation extends over the components $m$ in the $i$-th partition. The scaling exponent $\overline{\tau}_q$ is given by $\sum\limits_i \widetilde{p}(i)^q \sim l^{\overline{\large \tau}_q}$ where $l=d/M$. Figure \ref{fig:fig1}(d) shows the scaling of $\overline{\tau}_q$ with $q$. The localized states, as is expected, have very feeble multifractal behavior. The fractal dimension $D_2$ has higher values for states which are more delocalized. These delocalized states also exhibit a faster approximately linear fall of the scaling exponent $\overline{\tau}_q$ with $q$. 

\begin{figure*}[ht]
\includegraphics[height=5.5cm,width=15cm]{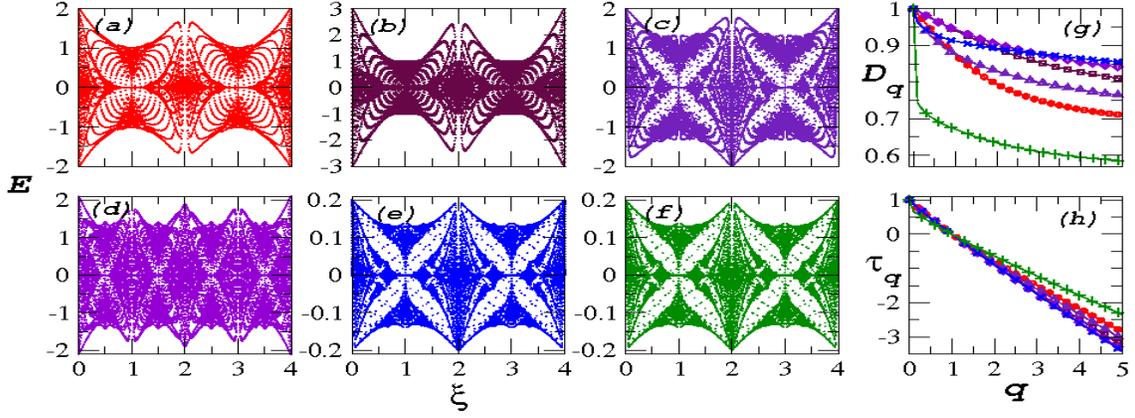}
\caption{(Color online) (a)-(f): The butterfly spectrum for the Hamiltonians summarized in Table \ref{table:tab1}. (g): Shows the multifractal property of these spectra.}
\label{fig:fig3}
\end{figure*}     

To study the statistical properties of the eigenstates, we consider the distribution of various quantifiers of fractal behavior. Fig. \ref{fig:fig2}(a)-(b) show the distribution of $\overline{D}_2=-\overline{\tau}_2$ and $\overline{D}_5 = -\overline{\tau}_5/4$ over all the eigenstates. We find that about $\sim 40\%$ of the eigenstates have very small values of the fractal dimensions. These states do not exhibit any fractal nature. Significant fractal behavior ($0.1 \lesssim (\overline{D}_2\, {\rm or} \, \overline{D}_5) \lesssim 0.8$) is exhibited by relatively small fraction of the eigenstates. Fig. \ref{fig:fig2}(c) shows the distribution of the slopes $\mu$ of the $\overline{\tau}_q\, {\rm vs.}\, q$ curves. We find that the large number of eigenstates which correspond to small values of $\overline{D}_2$ and $\overline{D}_5$ also do not show multifractality and for these states $\mu \sim 0$. We conclude that the eigenstates which exhibit fractal properties are small in number and these states also exhibit multifractality. The participation ratio (PR) for the $n$-th eigenstate defined as $PR=1/\sum_m |c_m^{(n)}|^4$ measures the number of basis states over which the given state has significant support. Fig. \ref{fig:fig2}(d) shows the distribution of the PR for all the eigenstates. We see that about $50\%$ of the eigenstates have PR less than $20$ which means that, for these states, out of $d=2001$ basis states, about $1980$ basis states do not have any component. These states are hence extremely localized. An important feature of the eigenstates of the effective Hamiltonian is the existence of a dominant proportion of localized states. These are also the states which do not exhibit fractal property.   

Having studied the self-similar spectrum of the effective Hamiltonian in Eq. (\ref{eq:heff_dkt}), we propose a general form of such Hamiltonians constructed using $SU(2)$ operators. We consider a Hamiltonian of the form
\be
\hh = a \jj_x + b \widehat{\mathbb{A}} + \left[ \widehat{C} \cos(\widehat{X}) + {\rm h.c.}\right]
\label{eq:gen_ham}
\ee
where $\widehat{\mathbb{A}} = \sum\limits_{m=-j}^{+j} \left(|m\rangle \langle m+1| + |m+1\rangle \langle m|\right)$ in the $\jj_z$ eigenbasis $\{|m\rangle\}$ and $\widehat{X} = \eta(2\jj_z+\mathbbm{1})/2j$. We note that the operator $\widehat{\mathbb{A}}$ is tri-diagonal in this representation with diagonal elements zero and off-diagonal elements are unity. The presence of the cosine term may lead to fractal spectrum for irrational values of $\eta/j$. Fig. \ref{fig:fig3} shows the energy spectrum for several choices of parameters $a, b$ and operator $\widehat{C}$ as summarized in Table \ref{table:tab1}. 

\begin{table}[h]
\begin{ruledtabular}
\begin{tabular}{ccccc}
~~Fig. &$~~a$ & $~~b$ & $~~\widehat{C} $ & $\mu~~$\\
\hline\hline
~~3(a)&$~~\alpha$ & $~~0$ & $~~\frac{\alpha}{2} (\jj_x + i \jj_y)$ & $ -0.697~~~$\\
~~3(b)&$~~\alpha$ & $~~0$ & $~~\alpha \jj_x$ & $-0.800~~$\\
~~3(c)&$~~\alpha$ & $~~0$ & $~~\frac{1}{2}\mathbbm{1}$ & $-0.756~~$\\
~~3(d)&$~~0$ & $~~\alpha$ & $~~\alpha \jj_x$ & $-0.833~~$\\
~~3(e)&$~~\alpha$ & $~~\epsilon\alpha$ & $~~\alpha \mathbbm{1}$ & $-0.851~~$\\
~~3(f)&$~~0$ & $~~\alpha$ & $~~\alpha \mathbbm{1}$ & $-0.579~~$\\
\end{tabular}
\end{ruledtabular}
\caption{}
\label{table:tab1}
\end{table}

All these cases reveal different forms of butterfly spectrum. Fig. \ref{fig:fig3}(a) is in fact very similar to the effective Hamiltonian for the double kicked top given in Eq. (\ref{eq:heff_dkt}). We note that Fig. \ref{fig:fig3}(a), (b) and (d) correspond to the Hamiltonian having only non-zero super- and sub-diagonal elements whereas \ref{fig:fig3}(c), (e), and (f) have both diagonal and off-diagonal elements. Fig. \ref{fig:fig3}(f) corresponds to the Harper/Aubry-Andre Hamiltonian where the super- and sub-diagonal elements are unity and the cosine modulation is along the diagonal \cite{harper,hoftstadter,aubry-andre}. Multifractal property of all these energy spectra is shown in Fig. \ref{fig:fig3}(g)-(h). Table \ref{table:tab1} summarizes the values of the slope $\mu$ for all the $\tau_q-q$ curves in Fig. \ref{fig:fig3}(h). Though the spectra look visually different from each other their fractal properties show remarkable similarity as is quantitatively encapsulated in Fig. \ref{fig:fig3}(h).

The Hamiltonian in Eq. (\ref{eq:gen_ham}) comprising of $SU(2)$ generators with $a=0, b=\alpha$ and $\widehat{C} = \alpha\mathbbm{1}$ has a close resemblance with the Hamiltonian representing non-interacting electrons moving in a $2D$ periodic square lattice with an external magnetic field. This variant of the usual Landau level problem where translation invariance of the lattice is broken in the presence of a constant magnetic field maps to the celebrated Harper/Aubry-Andre equation in the tight-binding approximation \cite{harper,hoftstadter,aubry-andre}. The Hamiltonian consists of an uniform nearest-neighbor hopping contribution and an onsite potential varying periodically with lattice site:
\be
\hh^{(h)} = \sum_{n=1}^L 2 \cos\bigl(2\pi n\sigma\bigr)\, |n\rangle \langle n| + \bigl(|n\rangle \langle n+1| \,+\, {\rm h.c.} \bigr)
\ee
where the summation extends over all the lattice sites. The corresponding energy spectrum for irrational $\sigma$ gives the Hoftstadter butterfly. If the onsite term is switched on and off at regular interval of time $T$ one has the kicked Harper model. The Floquet analysis for the kicked Harper model has been studied \cite{khm}. We investigate the effective time-independent approximate Hamiltonian obtained using Eq. (\ref{eq:heff_dkt}). We have assumed $\hht_0 = \sum\limits_{n=1}^L \alpha \bigl(|n\rangle \langle n+1| \,+\, {\rm h.c.} \bigr)$ and $\vvt = \sum\limits_{n=1}^L 2 \alpha \cos\bigl(2\pi n\sigma\bigr)\, |n\rangle \langle n|$ which yields
$\hh_{\rm eff}^{(h)} = \hh^{(h)} + \hh^{(corr)}$ where 
\be
\hh^{(corr)}= - \frac{1}{6} \sum_{n=1}^L\cos^2\bigl(2\pi n\sigma\bigr) \bigl(|n\rangle \langle n+1| \,+\, {\rm h.c.} \bigr).
\label{eq:heff_har}
\ee
The correction term in Eq. (\ref{eq:heff_har}) accounts for the kicked nature of the onsite potential in the effective static approximation.

\begin{figure}[b]
\includegraphics[height=7cm,width=8cm]{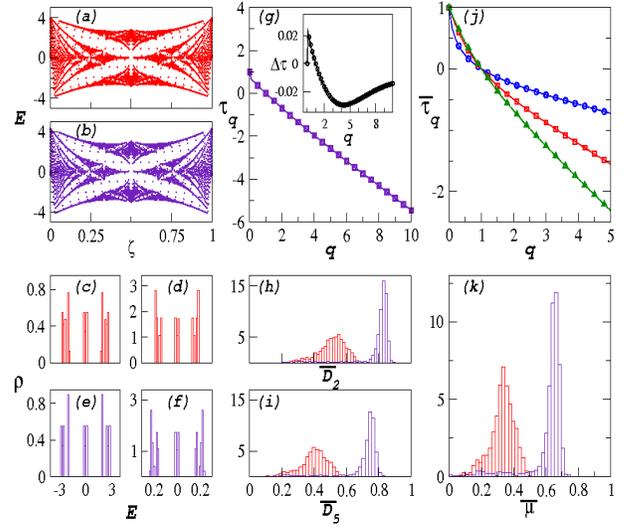}
\caption{(Color online) (a) Eigenspectrum of Harper Hamiltonian $\hh^{(h)}$. (b) Spectrum of the effective Hamiltonian for kicked Harper model $\hh_{\rm eff}^{(h)}$. (c)-(d) and (e)-(f) The spectral density at two different energy-scales revealing self-similarity for $\hh^{(h)}$ and $\hh_{\rm eff}^{(h)}$, respectively. (g) Multifractal behavior of the eigenvalues of $\hh_{\rm eff}^{(h)}$. Inset shows the deviation from $\hh^{(h)}$. (h)-(i) The distribution of $\overline{D}_2$ and $\overline{D}_5$ for the eigenvectors. In each of (h) and (i), the flat distribution in the left corresponds to $\hh^{(h)}$, and the sharp distribution on the right corresponds to $\hh_{\rm eff}^{(h)}$. (j) The multifractal property of few typical eigenvectors with $D_2$ values $0.248\, (\circ)$, $0.497\, (\square)$, and $0.684\, (\triangle)$. (k) The distribution of slope $\overline{\mu}$ of the $\overline{\tau}_q-q$ curves in the linear regime for all the eigenstates. The steeper distribution on the right of the figure corresponds to $\hh_{\rm eff}^{(h)}$ and the one to the left corresponds to $\hh^{(h)}$.} 
\label{fig:fig4}
\end{figure}

Figure \ref{fig:fig4}(a) and (b) show energy spectrum of $\hh^{(h)}$ and $\hh_{\rm eff}^{(h)}$, respectively. Both show almost similar butterfly pattern. The self-similar nature of the spectra for $\sigma = G_r$ is evident in Fig. \ref{fig:fig4}(c)-(f) which shows the spectral density at different energy scales. The multifractal nature of the spectrum of $\hh_{\rm eff}^{(h)}$ is shown in Fig. \ref{fig:fig4}(g). The linear fall of $\tau_q$ with $q$ has a slope of $-0.597$. The difference of $\tau_q$ for the spectrum of $\hh^{(h)}$ and $\hh_{\rm eff}^{(h)}$ as shown in the inset indicates that the multifractal properties of the eigenvalues are not significantly different for these Hamiltonians, with $|\Delta\tau|_{\rm max} \sim 0.02$. The fractal properties of the eigenvectors however show remarkable difference. Figure \ref{fig:fig4}(h) and (i) show the distribution for $\overline{D}_2$ and $\overline{D}_5$ for the eigenvectors of the two Hamiltonians. The mean $\overline{D}$ values for the eigenvectors are significantly different. The eigenstates of the $\hh_{\rm eff}^{(h)}$ show less variance around the mean fractal dimension as compared to the eigenstates of $\hh^{(h)}$. Figure \ref{fig:fig4}(j) shows the multifractal behaviors of three typical eigenvectors of $\hh_{\rm eff}^{(h)}$. The distribution of the slopes $\overline{\mu}$ of the $\overline{\tau}_q-q$ curves are compared for the eigenstates of $\hh^{(h)}$ and $\hh_{\rm eff}^{(h)}$ in Fig. \ref{fig:fig4}(k). Hence we find that the effective static Hamiltonian for the kicked Harper model though gives a very similar energy spectrum as the original Harper system, the fractal properties of the eigenvectors are considerably different for the two cases.

We conclude by noting that a wide class of kicked systems with Floquet butterfly spectrum also show self-similar behavior in the energy spectrum of their corresponding effective static Hamiltonian. The multifractality in the energy eigenstates are also found to be an useful quantifier to distinguish the self-similar properties of the spectrum for different Hamiltonians even when their eigenvalues show insignificant difference in fractal property. The effective Hamiltonians, though approximate, can be used to study statistical properties of such self-similar spectra for a wide range of time-dependent problems. 

\section{Supplementary material: Self-similar spectrum in effective time independent Hamiltonians for kicked systems}

We consider a general time-dependent Hamiltonian $\hh(t) = \hh_0 + \vv(t)$, with a time-periodic potential $\vv(t)=\vv(t+T)$ of periodicity $T$ has a Floquet operator $\fl(t)$ which is the time-evolution operator for one time-period. The method used in Refs. \cite{fishman,dalibard} factors the time-evolution unitary operator $\uu(t_i \rightarrow t_f)$ between times $t_i$ and $t_f = t_i + T$, as a sequence consisting of an initial kick followed by an evolution under a time-independent Hamiltonian and final kick \cite{dalibard}.
\be 
\uu(t_i \rightarrow t_f) =
\uue^\dagger(t_f) e^{-i \hh_{\rm eff} T}\, \uue(t_i)
\label{eq:floquet}
\ee
where $\uue(t) = e^{i \ff(t)}$ so that $\ff(t) = \ff(t+T)$ with vanishing average over one time period. For high-frequency forcing, the operators $\hh_{\rm eff}$ and $\ff(t)$ are expanded as a perturbation series in $1/\omega$ given by 
\be
\hh_{\rm eff} = \sum\limits_{n = 0}^\infty\, \frac{1}{\omega^n} \hh_{\rm eff}^{(n)},~~~~\ff(t) = \sum\limits_{n = 1}^\infty\, \frac{1}{\omega^n} \ff^{(n)}.
\ee 
This along with Eq. \eqref{eq:floquet} can be used to obtain 
$\hh_{\rm eff}$ and $\ff(t)$ up to any desired order of perturbation. At each order of perturbation, the average time-independent part, in this method, is retained in $\hh_{\rm eff}$ and all the time-dependence pushed to the operator $\ff(t)$. The convergence of the perturbation series is to be checked on case by case basis \cite{fishman,dalibard}. Expanding the periodic potential $\vv(t)$ in a Fourier series we have
\be 
\vv(t) = \vv_0 + \sum\limits_{n=1}^\infty \, \Bigl(\vv_{n} e^{i n \omega t} + \vv_{-n} e^{- i n \omega t}\Bigr).  
\ee
In terms of the Fourier coefficients, the truncated series for $\hh_{\rm eff}$ and $\ff(t)$ up to $\mathcal{O}(1/\omega^2)$ may be written as 
\begin{widetext}
\be
\begin{split}
\hh_{\rm eff} &= \hh_0 + \vv_0 + \frac{1}{\omega}
\sum\limits_{n=1}^\infty \frac{1}{n}\bigl[\vv_n, \vv_{-n}\bigr] +
\frac{1}{2\omega^2}\sum\limits_{n=1}^\infty \frac{1}{n^2} \Bigl(
\bigl[\bigl[ \vv_n, \hh_0\bigr], \vv_{-n}\bigr] + {\rm h.c.}\Bigr)\\
& + \frac{1}{3\omega^2} \sum\limits_{n, m = 1}^\infty \frac{1}{nm}
\Bigl(\bigl[\vv_n, \bigl[\vv_m, \vv_{-n-m}\bigr]\bigr]\Bigr.  \Bigl.-
2 \bigl[\vv_n, \bigl[\vv_{-m}, \vv_{m-n}\bigr]\bigr] + {\rm
  h.c.}\Bigr) \\ \ff(t) &=
\frac{1}{i\omega}\displaystyle\sum_{n=1}^{\infty}\frac{1}{n}\Bigl(\vv_ne^{in\omega
  t}-\vv_{-n}e^{-in\omega t}\Bigr) +
\frac{1}{i\omega^2}\displaystyle\sum_{n=1}^{\infty}\frac{1}{n^2}\Bigl(\bigl[\vv_n,\hh_0+\vv_0
  \bigr]e^{in\omega t} - {\rm h.c.}\Bigr)\\ &+ \frac{1}{2i\omega^2}
\sum\limits_{n, m = 1}^\infty \frac{1}{n (n+m)}
\Bigl(\bigl[\vv_n,\vv_m] e^{i(n+m)\omega t} - {\rm h.c.}\Bigr)+
\frac{1}{2i\omega^2}\sum\limits_{n\ne m = 1}^\infty \frac{1}{n (n-m)}
\Bigl(\bigl[\vv_n,\vv_{-m}\bigr] e^{i(n-m)\omega t} - {\rm
  h.c.}\Bigr).
\end{split}
\label{eq:hamnf}
\ee 
\end{widetext}
This general expression for the approximate effective static Hamiltonian for periodically driven systems is used in the article.

\end{document}